\documentclass{article}
\usepackage{ijcai24}

\usepackage{times}
\usepackage{soul}
\usepackage{url}
\usepackage[hidelinks]{hyperref}
\usepackage[utf8]{inputenc}
\usepackage[small]{caption}
\usepackage{graphicx}
\usepackage{amsmath}
\usepackage{amsthm}
\usepackage{booktabs}
\usepackage{algorithm}
\usepackage{algorithmic}
\usepackage[switch]{lineno}
\usepackage{multirow}
\usepackage{makecell}
\usepackage{subcaption}
\usepackage{tikz}
\usepackage{float}
\usepackage{adjustbox}

\title{LSP Framework: A Compensatory Model for Defeating Trigger Reverse Engineering via Label Smoothing Poisoning}

\author{
Beichen Li$^1$
\and
Yuanfang Guo$^1$\and
Heqi Peng$^1$\and
Yangxi Li$^2$\and
Yunhong Wang$^1$\and
\affiliations
$^1$School of Computer Science and Engineering, Beihang University, China\\
$^2$National Computer network Emergency Response technical Team/Coordination Center of China\\
\emails
\{libeichen, andyguo, penghq\}@buaa.edu.cn,
liyangxi@outlook.com,
yhwang@buaa.edu.cn
}

\begin{document}

\maketitle

\begin{abstract}
    Deep neural networks are vulnerable to backdoor attacks. Among the existing backdoor defense methods, trigger reverse engineering based approaches, which reconstruct the backdoor triggers via optimizations, are the most versatile and effective ones compared to other types of methods. In this paper, we summarize and construct a generic paradigm for the typical trigger reverse engineering process. Based on this paradigm, we propose a new perspective to defeat trigger reverse engineering by manipulating the classification confidence of backdoor samples. To determine the specific modifications of classification confidence, we propose a compensatory model to compute the lower bound of the modification. With proper modifications, the backdoor attack can easily bypass the trigger reverse engineering based methods. To achieve this objective, we propose a Label Smoothing Poisoning (LSP) framework, which leverages label smoothing to specifically manipulate the classification confidences of backdoor samples. Extensive experiments demonstrate that the proposed work can defeat the state-of-the-art trigger reverse engineering based methods, and possess good compatibility with a variety of existing backdoor attacks.
\end{abstract}

\section{Introduction}

\begin{figure}[!t]
    \centering
    \includegraphics[width=1.0\linewidth]{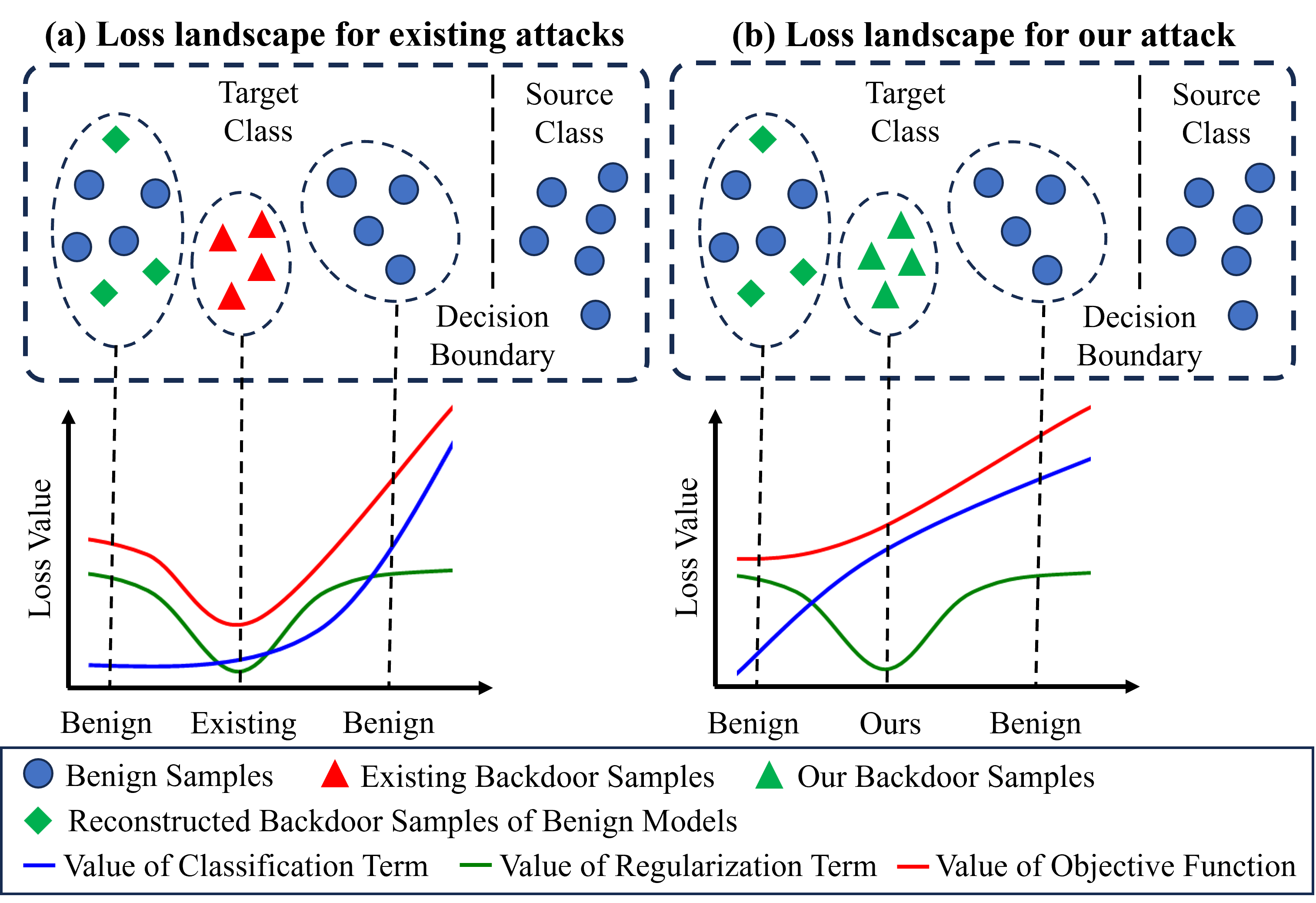}
    \caption{The illustration of our proposed work. Since the value of the classification and regularization terms are both very low, the value of the objective function of existing backdoor triggers is much smaller than that of universal adversarial patches, which makes the triggers easy to be reconstructed. Our work enlarges the value of classification term without changing the regularization term, which tends to change the value of objective function to be much larger than the minimum point.}
    \label{fig:introduction}
    \vspace{-0.3cm}
\end{figure}

Deep Neural Networks (DNNs) have shown great success in various scenarios, including many security crucial ones, such as autonomous driving \cite{hu2023planning} and face recognition \cite{marriott20213d}. Unfortunately, due to the requirements of huge computing resources and training data, an increasing number of users tend to outsource the DNN training tasks to professional AI companies, or directly download and employ the pretrained models from the internet. 

In such scenarios, an adversary can easily inject backdoors into a DNN model during the training process. When the backdoored model is deployed by the users, it will misclassify any data containing the backdoor trigger to a pre-defined target class. Existing researches have demonstrated that backdoor attacks can induce serious consequences in vital applications, such as autonomous driving systems \cite{DBLP:conf/ccs/LinXL020} and face recognition systems \cite{DBLP:conf/cvpr/WengerPBY0Z21}.

Various defense methods have been developed to address this security issue. These methods can be briefly categorized into trigger reverse engineering based \cite{DBLP:conf/sp/WangYSLVZZ19}, pruning based \cite{DBLP:conf/raid/0017DG18}, dataset inspecting based \cite{DBLP:conf/aaai/ChenCBLELMS19}, poisoned dataset utilization based \cite{DBLP:conf/iclr/Huang0WQ022} and preprocessing based \cite{DBLP:conf/asiaccs/0001ZGZQT21} methods.
Unfortunately, these methods, except for the trigger reverse engineering based methods, possess obvious weaknesses such as performance degradation on benign samples, requiring full access to the original poisoned dataset, or inability to handle less obvious triggers. 

On the contrary, trigger reverse engineering based methods (a.k.a. reversing based methods), such as Neural Cleanse \cite{DBLP:conf/sp/WangYSLVZZ19}, ABS \cite{DBLP:conf/ccs/LiuLTMAZ19} and ExRay \cite{DBLP:conf/cvpr/LiuSTWM022}, are considered the most versatile and effective.
These methods designs their objective functions based on certain assumption obtained by prior analysis of backdoor behaviors, to estimate and reconstruct certain backdoor triggers, which function similarly as the original triggers. Note that the reversing based methods also generate triggers from the benign models. However, since these reconstructed triggers do not satisfy their assumptions, the value of objective function is much large than that of the backdoored models.

In this paper, by analyzing the objective functions designed by the existing reversing based methods, we can obtain a generic paradigm that these objective functions consist of two major components. The first component (named the classification term) ensures that the reconstructed backdoor samples can be classified into the target class with high classification confidence. The second component (named the regularization term) constrains certain characteristics of the backdoor triggers (e.g., L1 norm of trigger patterns or SSIM of the reconstructed samples). By combining the two specific terms into one objective function, reversing based methods can identify the existence of backdoor triggers when the minimum of objective function is achieved.

To defeat the reversing based methods, existing backdoor attacks \cite{DBLP:conf/cvpr/Jiang0X023,DBLP:conf/cvpr/Zhao0XDWL22,DBLP:conf/iclr/NguyenT21,DBLP:journals/corr/abs-1912-02771,DBLP:journals/corr/abs-1712-05526} only focus on neutralize the regularization term, i.e., they usually propose various new backdoor triggers to challenge the assumption of previous reversing based methods.

Different from the existing backdoor attacks, in this paper, we propose a completely new, simple yet effective perspective to defeat the reversing based methods, i.e., we can manipulate the classification term of the generic paradigm to compensate for the variations in the regularization term, as shown in Fig. \ref{fig:introduction}. Specifically, when a reversing based method is employed for backdoor detection, the value of the regularization term of a backdoored model is unavoidably lower than that of the benign models. In our opinion, this discrepancy can be exploited as a characteristic of backdoor attack rather than a weakness. If we can manage to increase the value of the classification term to compensate for the decrease in the regularization term, which will deviate the backdoor trigger from the minimum point of the objective function, the reversing based methods cannot reconstruct the backdoor trigger correctly and their defense will then collapse.

To achieve this objective, the compensation to the regularization term is desired to be specifically estimated. Based on our generic paradigm, we construct a compensatory model, which desires to make the objective function value of the poisoned model larger than that of the benign model, to compute the lower bound of the compensation. Then, we can decrease the classification confidence of the backdoor samples to boost the value of the classification term, i.e. increase the compensation. If we force this increase to be equal to or larger than the lower bound of the compensation, the reversing based methods will then malfunction.

To utilize the above compensatory model to defeat the reversing based methods, we exploit label smoothing \cite{DBLP:conf/cvpr/SzegedyVISW16} to control the classification confidence of backdoor samples, and propose a simple yet effective Label Smoothing Poisoning (LSP) framework. Since most of the existing backdoor attacks does not rely on one-hot labels, our proposed LSP framework possesses excellent compatibility, i.e., it can be easily applied to the existing backdoor attack methods to significantly improve their attacking performance against the reversing based methods.

Our contributions are summarized as follows:

\begin{itemize}
    \item We construct a generic paradigm for the trigger reverse engineering based methods.
    \item We propose a new attack perspective to manipulate the classification term of the generic paradigm to compensate for the variations in the regularization term, to defeat the trigger reverse engineering based methods. To the best of our knowledge, we are the first work to reveal this type of design flaws in reversing based methods.
    \item We propose a compensatory model to compute the lower bound of the compensation to the regularization term and describe the relationship between the classification confidence of backdoor samples and the defense success rate of the trigger reverse engineering based methods.
    \item We propose the Label Smoothing Poisoning (LSP) framework to utilize the proposed compensatory model, which can be effectively integrated with the existing backdoor attacks.
\end{itemize}

\section{Related Work}

Backdoor attack is a certain type of adversarial attacks, which achieves the attack in the training or deploying process via data poisoning \cite{DBLP:journals/access/GuLDG19}, neural hijacking \cite{DBLP:conf/ndss/LiuMALZW018}, direct weight modification \cite{DBLP:conf/iclr/BaiWZL0X21} and etc. This paper focuses on studying the backdoor attacks in the training process, which inject malicious behaviors into models via modifying the training samples. In this section, we will respectively introduce the most widely used backdoor attacks and their most effective countermeasures.

\subsection{Backdoor Attack}

BadNets \cite{DBLP:journals/access/GuLDG19} pioneer to achieve a patch attack via stamping small square patches as backdoor triggers and relabeling them to the target class. After BadNets, most of the researches focus on improving the stealthiness of the attacks. \cite{DBLP:journals/corr/abs-1912-02771} improves BadNets to a clean label attack by only stamping triggers to the target class without relabeling. \cite{DBLP:journals/corr/abs-1712-05526} proposes a blend attack by adding consistent perturbations. \cite{DBLP:conf/cvpr/Jiang0X023} introduces color space transformations as backdoor triggers, which is the state-of-the-art filter attack. \cite{DBLP:conf/iclr/NguyenT21} and \cite{DBLP:conf/nips/NguyenT20} generates image-specific dynamic backdoor triggers to achieve a better stealthiness.

\subsection{Trigger Reverse Engineering}

Trigger reverse engineering based defense method is equivalent to computing an universal adversarial patch. This kind of methods requires a small clean dataset to estimate a consistent trigger pattern. Neural Cleanse \cite{DBLP:conf/sp/WangYSLVZZ19} is the first trigger reverse engineering based method. It assumes that the trigger patterns are desired to be small for stealthiness. Then, it constrains the L1 norm of the trigger patterns. It is effective against simple backdoor attacks such as BadNets. However, it fails to detect feature space attacks and dynamic attacks. ABS \cite{DBLP:conf/ccs/LiuLTMAZ19} assumes that a small subset of neurons, which are denoted as the compromised neurons, is correlated with the backdoor behaviour, and these neurons only activates if the trigger pattern exists. Thus, it imposes additional constraints on the activation values of compromised neurons. ExRay \cite{DBLP:conf/cvpr/LiuSTWM022} exploits the differences between the features of backdoor samples and benign samples to further check the validness of reconstructed triggers, which acts as a performance booster for other trigger reverse engineering based methods.

\section{Threat Model}

We consider backdoor attack and defense tasks in a MLaaS (Machine Learning as a Service) setting. The adversary possesses a benign dataset $X \in R^{W \times H \times C}$ and intends to train a backdoored model $F_{\omega}\left( x\right)$, which gives decent classification performance on benign samples $x$ and classifies the backdoor samples $x_{bd}$ to a predefined target class $y_t$. 
The backdoor samples $x_{bd}$ are obtained by applying a consistent trigger injecting function $f_{bd}\left( x\right)$ to the benign samples $x$. The defender is required to identify whether a model $F_{\omega}\left( x\right)$ is backdoored. The defender only has access to the benign dataset $X$ or other public datasets, while the attack target class $y_t$ and the trigger injecting function $f_{bd}\left( x\right)$ is unavailable.

\section{Proposed Work}

In this paper, we summarize a generic paradigm for the reversing based defense methods and defeat them from a new perspective, i.e., we manipulate the classification term of the generic paradigm to compensate for the decreases in the regularization term, which is induced by backdoor attacks. To estimate the specific compensation to the regularization term, we propose a compensatory model to compute the lower bound of the compensation. Based on the compensatory model, we propose a Label Smoothing Poisoning (LSP) framework by exploiting label smoothing \cite{DBLP:conf/cvpr/SzegedyVISW16} to control the classification confidence of the backdoor samples. 

\subsection{Reviews of the Existing Methods}\label{sec:review of existing methods}

We firstly review the reversing based methods. Due to limited space, we present two milestone methods.

\paragraph{Neural Cleanse.} 
By assuming that the triggers are small, Neural Cleanse \cite{DBLP:conf/sp/WangYSLVZZ19} reconstructs backdoor triggers, which are required to be classified to the target class, to simulate the potential backdoor samples. The objective function of Neural Cleanse is defined as 
\begin{eqnarray}
    \label{neural cleanse}
    &\min\limits_{m,\Delta}& L_{obj} = \ell_{ce}(y_t, F_{\omega}( x_{bd})) + \lambda \cdot \| m \|,
\end{eqnarray}
where $x_{bd} = f_{bd}(x)=(1-m)\cdot x+m \cdot \Delta$, $\forall x \in X$. Here, $\Delta$ denotes the trigger pattern. $m$ stands for the mask, which determines the size and position of the trigger to be added to the original image. $\Delta$ and $m$ constitute the trigger injecting function $f_{bd}(x)$, which is a simple mixing up operation. $\|\cdot\|$ is the L1 norm operator. $\ell_{ce}$ denotes the cross entropy loss of the reconstructed backdoor samples.

\paragraph{ABS.}
Besides of the constraints employed in Neural Cleanse, ABS \cite{DBLP:conf/ccs/LiuLTMAZ19} introduces a new assumption, i.e., the backdoored models can exhibit backdoor behavior, when a compromised neuron is activated while all the other neurons in the same layer remain inactive. In addition, ABS also exploits $\operatorname{SSIM}$ \cite{DBLP:journals/tip/WangBSS04} to maintain high visual quality in reconstructed samples.
The objective function of ABS can be summarized as below.
\begin{align}
    L_{obj} &= w_1\cdot \ell_{logits} + w_2\cdot \ell_{inter} + w_3\cdot \ell_{mask} \label{ABS} \\
    \ell_{logits} &= - F_\omega(x_{bd})[y_t] + \sum_{i\neq y_t} F_\omega(x_{bd})[i]\label{abs_loss_logits}\\
    \begin{split}
        \ell_{inter} &= -F_\omega(x_{bd}).layer[l].neuron[n] \\
        & + \sum_{i\neq n} F_\omega(x_{bd}).layer[l].neuron[i]
    \end{split}\label{abs_loss_inter}\\
    \ell_{mask} &= \operatorname{Max}(\| m \| - size, 0) + \operatorname{SSIM}(x,x_{bd})\label{abs_loss_mask}
\end{align}
Note that ABS constructs two trigger injecting functions, $x_{bd}=(1-m)\cdot x + m \cdot \Delta$ and $x_{bd}=\mathcal{F}(x)$ for the patch attack and filter attack settings, respectively, where $\mathcal{F}$ denotes a generative model. Here, $[\cdot]$ is the slice operator. $F_\omega(x_{bd})[i]$ stands for the logit value of the $i$-th class. $F_\omega(x_{bd}).layers[l].neuron[n]$ represents the activation value of the $n$-th neuron in the $l$-th layer. $size$ is a hyperparameter which determines the maximum norm of the reconstructed trigger. $\operatorname{SSIM}(x,x_{bd})$ computes the structure similarity index measure between benign samples and reconstructed backdoor samples.

\subsection{The Generic Paradigm}\label{sec:generic paradigm}

Based on our reviews, we can summarize that the objective functions of trigger reverse engineering based defense methods usually consist of two components, i.e., a classification term and a regularization term. Both of them are vital for trigger reverse engineering. Since the primary goal of a backdoor attack is to induce misclassification, reversing based methods are desired to ensure that their reconstructed triggers can be classified to the target class. Then, the classification term is vital.
Besides, since the universal adversarial patches (UAPs) \cite{DBLP:journals/corr/abs-1712-09665} can also induce misclassifications, which possess similar functionality as the backdoor triggers, the reversing based methods usually utilize a regularization term to force the reconstructed samples to be backdoor triggers rather than UAPs.

In general, the trigger reverse engineering based methods can be formulated as
\begin{equation}
\label{reversing paradigm}
\begin{aligned}
    &\min\limits_{x_{bd}} L_{obj} = \mathcal{L}_{cls}(x_{bd}, y_t) + \mathcal{R}( x_{bd}),
\end{aligned}
\end{equation}
where $x_{bd}$ is the reconstructed backdoor sample. $x_{bd}$ is generated by a trigger injecting function $f_{bd}(x)$ via $x_{bd} = f_{bd}(x)$. $\mathcal{L}_{cls}(x_{bd}, t)$ is the classification term to ensure that most of the reconstructed samples $x_{bd}$ are classified to the target class $y_t$. $\mathcal{R}(x_{bd})$ is a regularization term which guarantees that the reversed samples $x_{bd}$ satisfy specific prior assumptions. In the rest of this paper, we will abbreviate $\mathcal{L}_{cls}(x_{bd}, t)$ as $\mathcal{L}_{cls}$ and $\mathcal{R}(x_{bd})$ as $\mathcal{R}$ for simplicity.

\subsection{Compensatory Model}\label{sec:theory}

Since the reversing based methods treat the backdoor hunting process as finding the minimum value of $L_{obj}$, an intuitive countermeasure is to make the backdoor trigger no longer the minimum point. Many attack methods have attempted to defeat the reversing based methods through this perspective by violating their assumptions. In fact, this is equivalent to increase the value of $\mathcal{R}$. For example, Blend attack \cite{DBLP:journals/corr/abs-1712-05526} utilizes a watermark trigger, which spans the entire image, to enlarge $\lambda\cdot\| m \|$ in Eq. \eqref{neural cleanse} to defeat Neural Cleanse. This kind of attack perspective treats the reduction of $\mathcal{R}$ as a weakness of backdoor attacks. 

In this paper, we consider that the reduction of $\mathcal{R}$ is a characteristic of backdoor attack rather than a weakness, and thus propose a novel attack perspective, i.e., we can compensate the reduction of $\mathcal{R}$ by enlarging $\mathcal{L}_{cls}$. To achieve this objective, a question must be solved: \emph{"How much increase in $\mathcal{L}_{cls}$ would be necessary to undermine defense mechanisms?"}

Since the detection success rate of the reversing based methods is mainly related to the objective function values of the target class, in the following discussion, we only consider the target class. Let $L_{obj}^{ben}$ be the minimum value of $L_{obj}$ of the benign models, and $L_{obj}^{poi}$ be that of the backdoored models. In the generic paradigm, $L_{obj}^{poi}$ should be much lower than $L_{obj}^{ben}$ to ensure that the reverse engineering process can finally reach the backdoor trigger. Therefore, we can simply violate this assumption via
\begin{equation}
\label{base assumption}
    L_{obj}^{poi} \geq L_{obj}^{ben}.
\end{equation}

By substituting Eq. \eqref{reversing paradigm} into Eq. \eqref{base assumption}, we can obtain
\begin{equation}
\label{substituting}
    \mathcal{L}_{cls}^{poi} + \mathcal{R}^{poi} \geq \mathcal{L}_{cls}^{ben} + \mathcal{R}^{ben},
\end{equation}
where, $\mathcal{L}_{cls}^{poi}$, $\mathcal{R}^{poi}$, $\mathcal{L}_{cls}^{ben}$, $\mathcal{R}^{ben}$ denote the classification term of the backdoored models, the regularization term of backdoored models, the classification term of benign models, the regularization term of benign models, respectively.

Then, Eq. \eqref{substituting} can be reformed to 
\begin{equation}
\label{attack theory}
    \mathcal{L}_{cls}^{poi} \geq \mathcal{R}^{ben} - \mathcal{R}^{poi} + \mathcal{L}_{cls}^{ben}.
\end{equation}

Our compensatory model, which is described by Eqs. \eqref{substituting} and \eqref{attack theory}, actually reveals a design flaw in reversing based methods. Specifically, we can reduce the classification confidence of backdoor samples, $\mathcal{L}_{cls}^{poi}$, to compensate the decrease of $\mathcal{R}^{poi}$ compared to $\mathcal{R}^{ben}$, until the overall loss value of the poisoned model is equal to or larger than that of the benign model. As a result, the backdoor samples will no longer be the minimum point of the objective function and can successfully bypass the reversing based methods.

The specific values of $\mathcal{R}^{ben}$, $\mathcal{R}^{poi}$ and $\mathcal{L}_{cls}^{ben}$ are determined by the specific defense methods. For example, if we intend to defeat Neural Cleanse, we can substitute Eq. \eqref{neural cleanse} into Eq. \eqref{attack theory}. Since the benign models are required to possess high classification confidence on the reconstructed samples of Neural Cleanse, $\mathcal{L}_{cls}^{ben}$ is small and it can be omitted. At last, we can obtain
\begin{equation}
\label{attack neural cleanse}
    \ell_{ce} (y_t, F_\omega(x_{bd})) \geq \lambda (\| m^{ben} \| - \| m^{poi} \|) + \epsilon.
\end{equation}
Eq. \eqref{attack neural cleanse} gives the lower bound of the cross entropy loss to ensure an undetectable attack against Neural Cleanse. If we want to defeat the more complex methods such as ExRay, $\mathcal{R}^{ben}$, $\mathcal{R}^{poi}$ and $\mathcal{L}_{cls}^{ben}$ can be directly obtained via running these methods. 

In practice, we observe that the classification confidence for the target class is usually quite large, i.e., the potential space for the compensation is quite large. Since the classification confidence of the target class will not affect the final prediction result when it is above the classification threshold, we can still perform successful backdoor attacks at a high attack success rate even if we utilize a relatively large compensation. The experimental evidence for this hypothesis is  given in Sec. \ref{sec:attack rate}.

\begin{algorithm}[tb]
    \caption{Label Smoothing Poisoning Framework}
    \label{algo:LSP Attack}
    \textbf{Input}: Attack rate ${ar}$, trigger injecting function $f_{bd}\left( x\right)$ from the selected baseline attack, attack target class $y_t$, poisoning rate $r_p$, maximum number of training epochs $epochs$, clean dataset $X$ consist of (sample, label) pairs $\left( x,y\right)$\\
    \textbf{Output}: Backdoored model $M\left( x\right)$
    \begin{algorithmic}[1] 
        \STATE $num\_benign=len\left( X\right) \cdot \left( 1-r_p\right)$
        \STATE $X^{ben}=X[:num\_benign]$
        \STATE $X^{poi}=X[num\_benign:]$
        \STATE $count=0$
        \WHILE{$count<epochs$}
            \FOR{$\left( x,y\right)$ in $X^{ben} \cup X^{poi}$}
                \IF{$x \in X^{poi}$}
                    \STATE $x=f_{bd}\left( x\right)$
                    \STATE $y=constrained\_label\_smoothing\left( y_t,ar\right)$
                \ENDIF
                \STATE train $M$ with $\left( x,y\right)$ by cross entropy loss
            \ENDFOR
            \STATE $count=count+1$
        \ENDWHILE
        \STATE \textbf{return} $M$
    \end{algorithmic}
\end{algorithm}

\begin{table*}[tb]
    \vspace{-0.5cm}
    \small
    \centering
    \begin{tabular}{cccccccccc}
        \toprule
        \multirow{2}{*}{\makecell{Dataset\\(Model Architecture)}} & \multirow{2}{*}{Attack Method} & \multirow{2}{*}{BA} & \multirow{2}{*}{ASR} & \multicolumn{2}{c}{Neural Cleanse} & \multicolumn{2}{c}{ABS} & \multicolumn{2}{c}{ExRay} \\
        \cline{5-10}
        \rule{0pt}{11pt}
         & & & & ACC & AP & ACC & AP & ACC & AP \\
        \midrule
        \multirow{5}{*}{\makecell{FMNIST\\(VGG16)}} & Clean & 0.902 & \multicolumn{1}{c}{-} & - & - & - & - & - & - \\
         & BadNet       & 0.902 & 0.983 & 0.695 & 0.807 & 0.910 & 0.969 & 0.640 & 0.716 \\
         & BadNet-LSP   & 0.902 & 0.979 & 0.510 & 0.527 & 0.660 & 0.631 & 0.565 & 0.611 \\
         & LC           & 0.903 & 0.995 & 0.505 & 0.522 & 0.570 & 0.577 & 0.560 & 0.554 \\
         & LC-LSP       & 0.903 & 0.991 & 0.490 & 0.501 & 0.590 & 0.511 & 0.510 & 0.486 \\
         \midrule
         \multirow{5}{*}{\makecell{CIFAR10\\(ResNet18)}} & Clean & 0.924 & \multicolumn{1}{c}{-} & - & - & - & - & - & - \\
         & RandSQ       & 0.926 & 1.000 & 0.735 & 0.740 & 0.650 & 0.714 & 0.805 & 0.881 \\
         & RandSQ-LSP   & 0.927 & 0.999  & 0.510 & 0.491 & 0.605 & 0.639 & 0.655 & 0.669 \\
         & WaNet        & 0.922 & 0.919  & - & - & 0.500 & 0.438 & 0.510 & 0.456 \\
         & WaNet-LSP    & 0.935 & 0.925  & - & - & 0.500 & 0.465 & 0.500 & 0.375 \\
         \midrule
         \multirow{3}{*}{\makecell{GTSRB\\(ResNet18)}} & Clean & 0.955 & \multicolumn{1}{c}{-} & - & - & - & - & - & - \\
         & ColorBD        & 0.953 & 1.000 & - & - & 0.620 & 0.679 & 0.970 & 0.990 \\
         & ColorBD-LSP    & 0.953 & 0.994  & - & - & 0.560 & 0.551 & 0.565 & 0.598 \\
         \midrule
         \multirow{3}{*}{\makecell{ImageNet\\(WideResNet50)}} & Clean & 0.767 & \multicolumn{1}{c}{-} & - & - & - & - & - & - \\
         & Blend        & 0.765 & 1.000 & - & - & 0.880 & 0.929 & 0.740 & 0.825 \\
         & Blend-LSP    & 0.765 & 0.994  & - & - & 0.555 & 0.585 & 0.535 & 0.531 \\
        \bottomrule
    \end{tabular}
    \caption{Results of the baseline attack methods and our proposed LSP framework. Since WaNet, ColorBD and Blend are not patch attack, we have not evaluated them against Neural Cleanse.}
    \label{tab:main comparison}
    \vspace{-0.1cm}
\end{table*}

\begin{table}[tb]
    \small
    \centering
    \begin{tabular}{cccc}
        \toprule
        Dataset & Model Architecture & \#Classes & \#Benign \\
        \midrule
        FMNIST & VGG16 & 10 & 40000 \\
        CIFAR10 & ResNet18 & 10 & 40000 \\
        GTSRB & ResNet18 & 43 & 39209 \\
        ImageNet & WideResNet50 & 200 & 80000\\
        \bottomrule
    \end{tabular}
    \caption{Datasets and model architectures.}
    \label{tab:datasets}
\end{table}

\begin{table}[tb]
    \small
    \centering
    \begin{tabular}{cccc}
        \toprule
        Method & Dataset & Poisoning Rate & Trojan Size \\
        \midrule
        BadNets & FMNIST    & 10\% & $16/\left( 28 \times 28 \right)$ \\
        LC      & FMNIST    & 50\% & $16/\left( 28 \times 28 \right)$ \\
        RandSQ  & CIFAR10   & 10\% & $9/\left( 32 \times 32 \right)$ \\
        WaNet   & CIFAR10   & 10\% & Dynamic \\
        ColorBD & GTSRB     & 5\% & Whole Image \\
        Blend   & ImageNet  & 10\% & Whole Image \\
        \bottomrule
    \end{tabular}
    \caption{Details of backdoor attacks.}
    \label{tab:attack details}
\end{table}

\subsection{Label Smoothing Poisoning Framework}\label{sec:framework}

Sec. \ref{sec:theory} has demonstrated that the key of bypassing the reversing based methods is to reduce the classification confidence of backdoor samples to compensate the decreases in the regularization term. Besides, it is also important to maintain the classification confidence of target class to be the highest to ensure a successful backdoor attack. Then, we adopt label smoothing \cite{DBLP:conf/cvpr/SzegedyVISW16} to construct our constrained label smoothing function, to simultaneously achieve the above two intentions, as 

\begin{align}
    \label{label convert}
    logits_i &=\left\{\begin{matrix}
    ar,&i=y_t\\
    1,&\text{otherwise}
    \end{matrix}\right.\\
    \label{label softmax}
    label& = \operatorname{softmax} \left ( logits \right ).
\end{align}

Here, $logits$ is an array of logits, and $logits_i$ represents the logit value of the $i$-th class. $y_t$ stands for the target class defined by the adversary. $ar$, which is the attack rate, is a hyper parameter to control the attack strength. Note that $ar$ should be greater than 1 to ensure a successful backdoor attack. When $ar=+\infty$, our LSP framework will degenerate to the traditional backdoor attacks. Considering that most of the poisoning attacks employs the cross entropy loss as supervision, we convert $logits$ to ground truth labels by $\operatorname{softmax}$. 

Algorithm \ref{algo:LSP Attack} describes the detailed procedures of our LSP framework. Since most of the existing poisoning attacks does not require modifications of ground truth labels, our framework is compatible to all these attacks by simply loading the specific trigger injecting function $f_{bd}\left( x\right)$.

\section{Experiment}

\subsection{Experiment Setup}\label{sec:exp setup}

\paragraph{Datasets.} 
By following the existing literatures \cite{DBLP:conf/sp/WangYSLVZZ19,DBLP:conf/ccs/LiuLTMAZ19,DBLP:conf/cvpr/LiuSTWM022}, FMNIST \cite{DBLP:journals/corr/abs-1708-07747}, CIFAR10 \cite{DBLP:journals/corr/abs-1811-07270} and GTSRB \cite{DBLP:journals/nn/StallkampSSI12} are employed as datasets. Tab. \ref{tab:datasets} provides an overview of these datasets. To demonstrate the effectiveness of our LSP framework in more challenging scenarios, we also employ TinyImageNet \cite{le2015tiny}. 
To prevent the overlap between the poisoned and benign data, we equally divide the data from each class into two set. One set is employed to generate the poisoned samples, and the other servers as benign training dataset.

\paragraph{Baseline backdoor attacks.} We select $5$ attack baseline methods with different attacking mechanisms, i.e., BadNets \cite{DBLP:journals/access/GuLDG19}, Label-Consistent Attack (abbreviated as LC) \cite{DBLP:journals/corr/abs-1912-02771}, Blend Attack \cite{DBLP:journals/corr/abs-1712-05526}, WaNet \cite{DBLP:conf/iclr/NguyenT21} and Color Backdoor (abbreviated as ColorBD) \cite{DBLP:conf/cvpr/Jiang0X023}. To further enhance the diversity of triggers, we implement RandSQ, which introduces randomness to the positions and colors of the triggers designed by BadNets. Details of these attacks are given in Tab. \ref{tab:attack details}. Examples of these attacks are given in the supplementary file.

\paragraph{Backdoored and benign models.} We selected the classes 0 to 9 as the target labels. each of these target labels is trained 10 times with different random seeds, to obtain 100 backdoored models for each backdoor attack. Correspondingly, we employ the same setting to train 100 backdoored models for each baseline attack with our LSP framework applied, via Algorithm \ref{algo:LSP Attack}. The attack rate for BadNets, LC, RandSQ, WaNet, ColorBD, Blend are 3.5, 3.7, 2.0, 2.3, 1.6, 2.0, respectively, which are computed by our compensatory model defined by Eq. \eqref{attack theory}. The only difference between a baseline attack and its LSP version is that the LSP version uses the constrained label smoothing function to change the ground truth labels of poisoned samples. If not explicitly stated, the value of attack rate $ar$ is defaultly set to 2.0. The number of baseline attacked models, LSP attacked models and benign models are 600, 600 and 400, respectively.

\paragraph{Backdoor defense methods.} 
Three reversing based methods, i.e., Neural Cleanse \cite{DBLP:conf/sp/WangYSLVZZ19}, ABS \cite{DBLP:conf/ccs/LiuLTMAZ19} and ExRay \cite{DBLP:conf/cvpr/LiuSTWM022}, are employed to evaluate our method. Due to the unavailability of the PyTorch version for Neural Cleanse, it is re-implemented by referring to the official implementation in Keras. For ABS and ExRay, their official implementations on GitHub are directly employed.

\paragraph{Evaluation metrics.} For the attack task, we employ benign accuracy (denoted as BA) and attack success rate (denoted as ASR) as the evaluation metrics. For the backdoor detection task, we utilize defense accuracy (denoted as ACC) and average precision (denoted as AP) to measure the effectiveness of reversing based methods.

\begin{table*}[tb]
    \vspace{-0.5cm}
    \small
    \centering
    \subcaptionbox{BadNets}{
        \begin{tabular}{ccccccccccc}
            \toprule
            \multirow{2}{*}{Target} & \multicolumn{10}{c}{Averaged Norm} \\
            \cline{2-11}
            \rule{0pt}{10pt}
            & 0 & 1 & 2 & 3 & 4 & 5 & 6 & 7 & 8 & 9 \\
            \midrule
            Ben & 41 & 62 & 38 & 62 & 48 & 45 & 38 & 68 & 26 & 72 \\
            \midrule
            0 & \textbf{\underline{10}} & 57 & 37 & 60 & 45 & 48 & 35 & 66 & 26 & 72 \\
            1 & 38 & \textbf{\underline{9}} & 39 & 57 & 46 & 48 & 38 & 69 & 27 & 76 \\
            2 & 38 & 56 & \textbf{\underline{9}} & 60 & 45 & 47 & 37 & 67 & 26 & 71 \\
            3 & 37 & 52 & 38 & \textbf{\underline{10}} & 45 & 45 & 35 & 65 & 27 & 70 \\
            4 & 36 & 58 & 35 & 59 & \textbf{\underline{10}} & 50 & 37 & 67 & 27 & 77 \\
            5 & 38 & 60 & 41 & 64 & 45 & \textbf{\underline{9}} & 37 & 63 & 26 & 69 \\
            6 & 39 & 56 & 40 & 62 & 46 & 47 & \textbf{\underline{10}} & 63 & 26 & 78 \\
            7 & 36 & 58 & 38 & 61 & 45 & 46 & 36 & \textbf{\underline{9}} & 25 & 69 \\
            8 & 38 & 58 & 41 & 66 & 50 & 47 & 37 & 68 & \textbf{\underline{8}} & 78 \\
            9 & 39 & 63 & 41 & 61 & 47 & 45 & 37 & 61 & 26 & \textbf{\underline{9}} \\
            \bottomrule
        \end{tabular}
    }
    \subcaptionbox{BadNets-LSP}{
        \begin{tabular}{ccccccccccc}
            \toprule
            \multirow{2}{*}{Target} & \multicolumn{10}{c}{Averaged Norm} \\
            \cline{2-11}
            \rule{0pt}{10pt}
            & 0 & 1 & 2 & 3 & 4 & 5 & 6 & 7 & 8 & 9 \\
            \midrule
            Ben & 41 & 62 & 38 & 62 & 48 & 45 & 38 & 68 & 26 & 72 \\
            \midrule
            0 & \underline{38} & 61 & 36 & 60 & 43 & 46 & 35 & 61 & \textbf{23} & 67 \\
            1 & 38 & \underline{61} & 36 & 58 & 44 & 44 & 34 & 68 & \textbf{24} & 67 \\
            2 & 39 & 59 & \underline{33} & 58 & 43 & 45 & 34 & 58 & \textbf{23} & 68 \\
            3 & 36 & 54 & 35 & \underline{57} & 42 & 44 & 35 & 57 & \textbf{22} & 64 \\
            4 & 35 & 59 & 37 & 60 & \underline{42} & 45 & 34 & 60 & \textbf{23} & 68 \\
            5 & 43 & 60 & 35 & 58 & 45 & \underline{44} & 35 & 59 & \textbf{24} & 70 \\
            6 & 36 & 56 & 37 & 61 & 42 & 43 & \underline{34} & 55 & \textbf{23} & 67 \\
            7 & 36 & 57 & 36 & 57 & 43 & 45 & 36 & \underline{58} & \textbf{23} & 66 \\
            8 & 36 & 58 & 35 & 59 & 44 & 43 & 35 & 58 & \textbf{\underline{22}} & 66 \\
            9 & 37 & 58 & 37 & 59 & 45 & 47 & 36 & 57 & \textbf{23} & \underline{66} \\
            \bottomrule
        \end{tabular}
    }
    \caption{Averaged norms in Neural Cleanse against BadNets. (a) Results against baseline BadNets. (b) Results against our LSP version. The minimum value of each target class is highlighted in bold letters, and the norms of ground truth target class is underlined.}
    \label{tab:NC-details}
    \vspace{-0.3cm}
\end{table*}

\subsection{Performance Evaluations}\label{sec:performance}

Tab. \ref{tab:main comparison} presents the attack results against the benign models as well as the three SOTA reversing based defense methods. Compared to the baseline attacks, since our LSP framework reduces the classification confidence of the backdoor samples, we gives a tiny performance drop in term of ASR against the benign models. Since the decreases in ASR are within 1\%, we believe it is negligible compared to the overall 95+\% ASR value. 
For the attack performances against the reversing based methods, it is obvious that all of the three defense methods give significant performance drops against the LSP versions of the attacks, compared to the baseline attacks. Even the simplest attack, BadNets, can bypass these defense methods when BadNets is integrated with our LSP framework.

\paragraph{Analysis on Neural Cleanse.} 
Tab. \ref{tab:NC-details} shows the averaged norms of triggers reversed by Neural Cleanse for each class. The `Ben' row represents the averaged norms of the benign models. The subsequent rows labeled 0-9 represent the averaged norms for backdoored models targeting at class 0-9, respectively. Then, three observations can be obtained: (1) Compared to the baseline attack, our LSP attack can give a significantly higher norm on the ground truth target classes. This indicates that the compensation to regularization term in our LSP framework hampers the trigger reverse engineering process on the target class. (2) When considering the norms of the ground truth target classes independently, we observe that they are nearly identical to the norms of their corresponding classes in the benign models. This implies that the reversed triggers on the ground truth target classes of the backdoored models can resemble similar trigger as that of the benign models. This illustrates that LSP framework can render the trigger to become undetectable to the reversing based methods. (3) Considering the overall trend of the norms, we observe a slight decrease in the overall norms with LSP framework. However, the overall trend is similar to that of the benign models, which indicates that our LSP framework does not significantly alter the models' feature space.

\begin{table}[tb]
    \small
    \centering
    \begin{tabular}{ccc}
        \toprule
        \multirow{2}{*}{Target Class} & \multicolumn{2}{c}{Attack Method} \\
        \cline{2-3}
        \rule{0pt}{10pt}
        & BadNets & BadNets-LSP \\
        \midrule
        0 & 0.945 & 0.526 \\
        1 & 0.913 & 0.931 \\
        2 & 0.995 & 0.505 \\
        3 & 0.904 & 0.413 \\
        4 & 0.960 & 0.441 \\
        5 & 0.996 & 0.893 \\
        6 & 1.000 & 0.714 \\
        7 & 0.977 & 0.461 \\
        8 & 0.985 & 0.718 \\
        9 & 0.979 & 0.638 \\
        \bottomrule
    \end{tabular}
    \caption{ABS scores of ground truth target class.}
    \label{tab:ABS-details}
\end{table}

\paragraph{Analysis on ABS.}
Tab. \ref{tab:ABS-details} shows the ABS scores of BadNets and its LSP version. As can be observed, most of the ABS scores for ground truth classes decrease obviously in LSP Attack. Since the results of ABS are similar to that of Neural Cleanse, due to space limitations, we will not delve into detailed analysis.

\subsection{Impact of Attack Rate}\label{sec:attack rate}

Here, BadNets is employed as the baseline attack because both Neural Cleanse and ABS perform decently against it.

\begin{figure}[tb]
    \centering
    
    \begin{subfigure}{0.8\linewidth}
        \centering
        \includegraphics[width=1\linewidth]{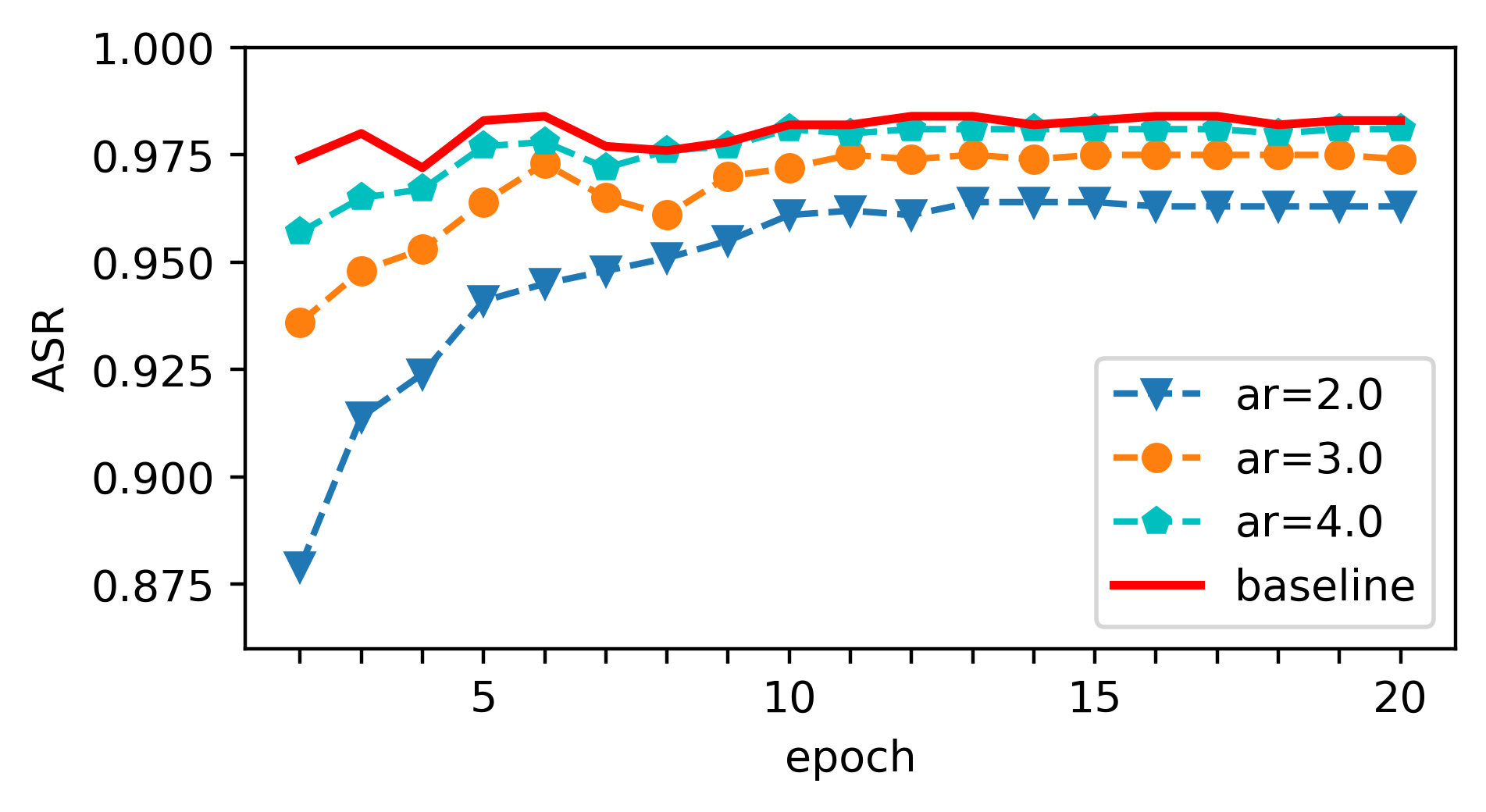}
        \captionsetup{skip=0pt}
        \caption{ASR v.s. attack rate}
        \label{fig:ASR-AR}
    \end{subfigure}

    \begin{subfigure}{0.8\linewidth}
        \centering
        \includegraphics[width=1\linewidth]{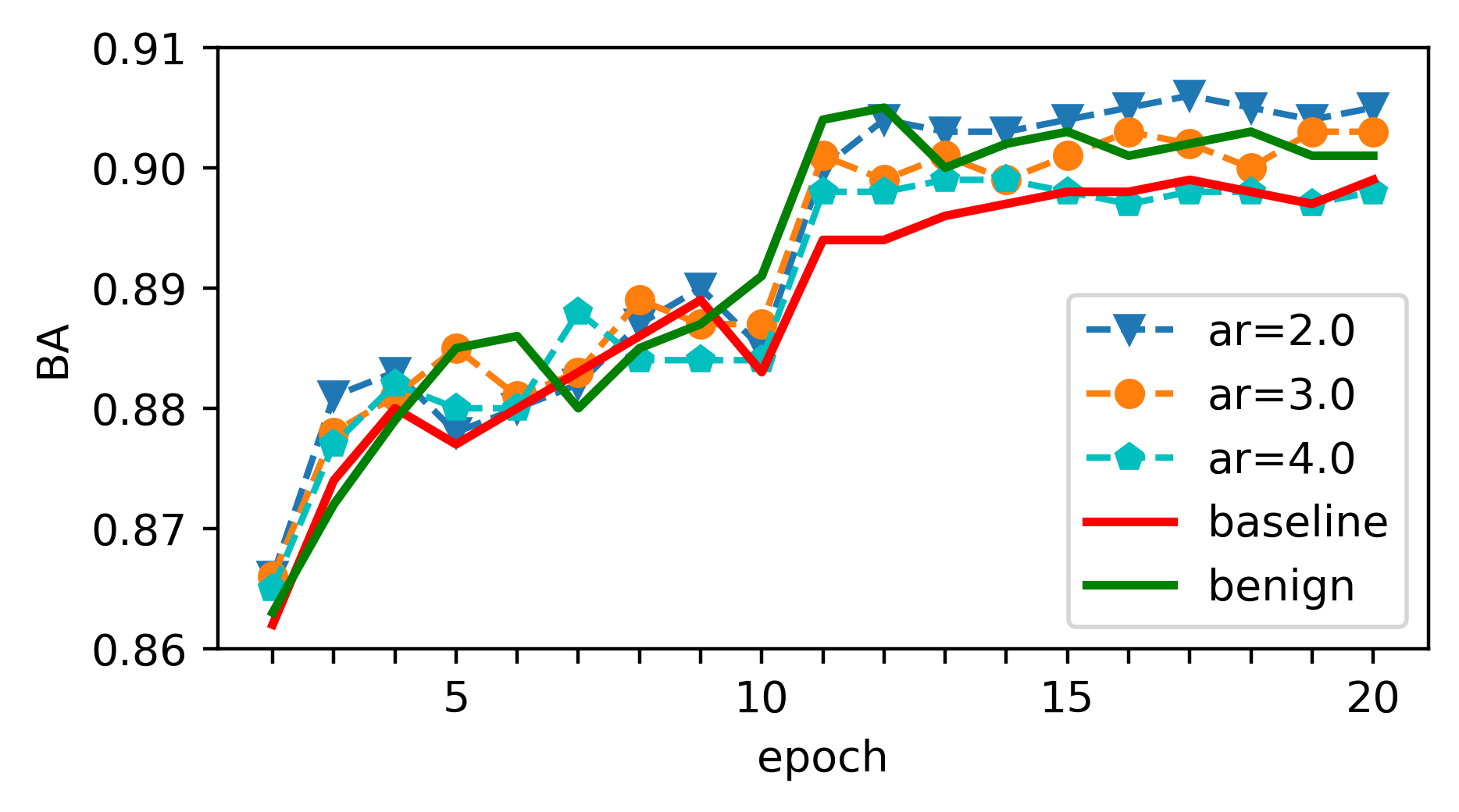}
        \captionsetup{skip=0pt}
        \caption{BA v.s. attack rate}
        \label{fig:BA-AR}
    \end{subfigure}
    \captionsetup{skip=1.3pt}
    \caption{The impacts of attack rate on ASR and BA. When the attack rate is 2.0, 3.0 and 4.0, the corresponding classification confidence on target class is 23.20\%, 45.09\% and 69.06\%, respectively.}
    \label{fig:attack rate variation}
    \vspace{-0.3cm}
\end{figure}

Fig. \ref{fig:attack rate variation} illustrates the relationship between ASR and attack rate. As can be observed from Fig. \ref{fig:ASR-AR}, LSP Attack converges slightly slower than the baseline. Still, both of them converges around 10 epochs. Fig. \ref{fig:BA-AR} indicates that the model's BA continues to improve even after 15 epochs. This suggests that though LSP Attack increases the difficulty of learning the backdoor data, model can still finish the learning of backdoor features before converging on the benign data. Therefore, our LSP framework will not demand additional training epochs to inject backdoors.

\begin{figure}[tb]
    \centering
    
    \begin{subfigure}{1\linewidth}
        \centering
        \includegraphics[width=0.9\linewidth]{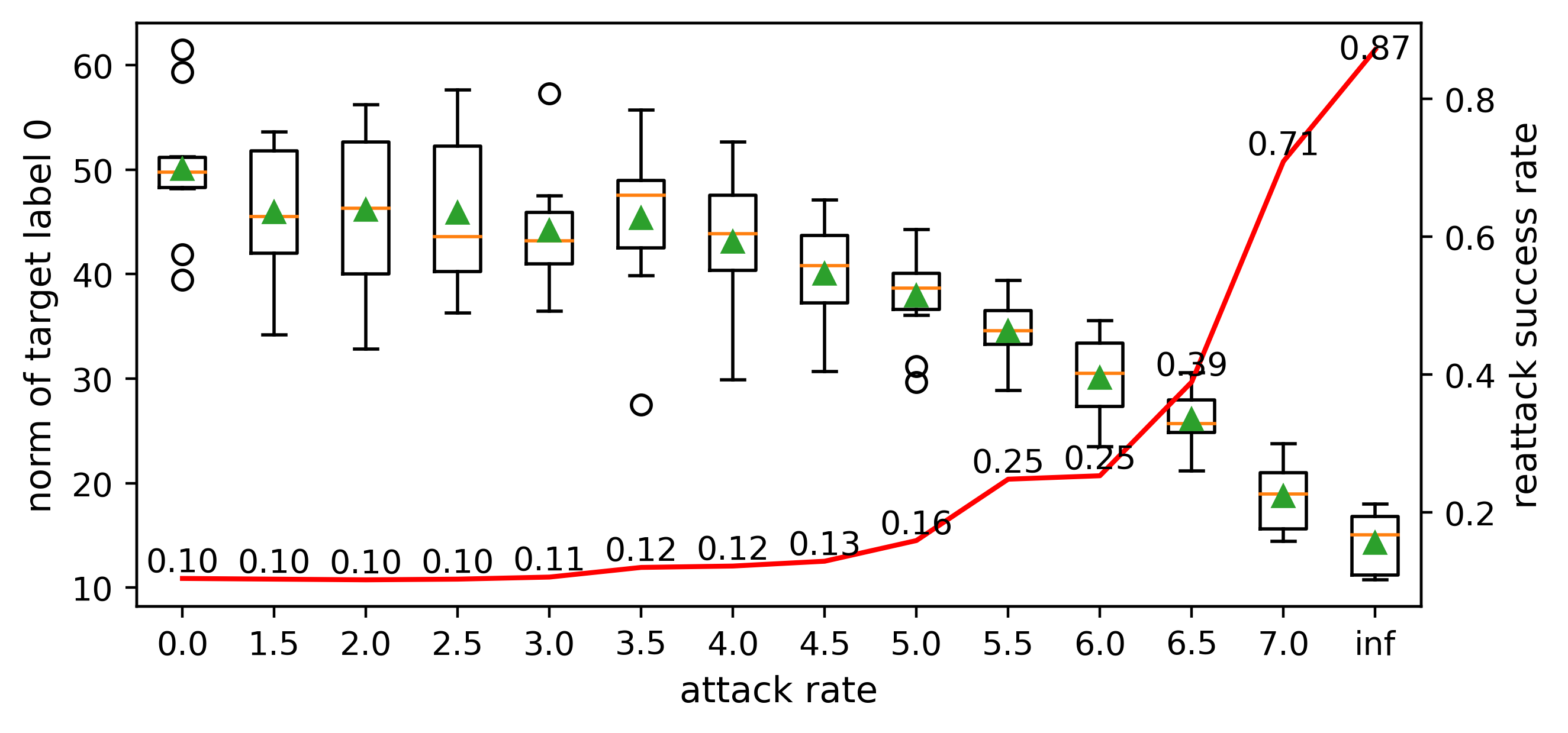}
        \captionsetup{skip=0pt}
        \caption{neural cleanse results (lambda=0.001)}
        \label{fig:nc-0.001}
    \end{subfigure}

    \begin{subfigure}{1\linewidth}
        \centering
        \includegraphics[width=0.9\linewidth]{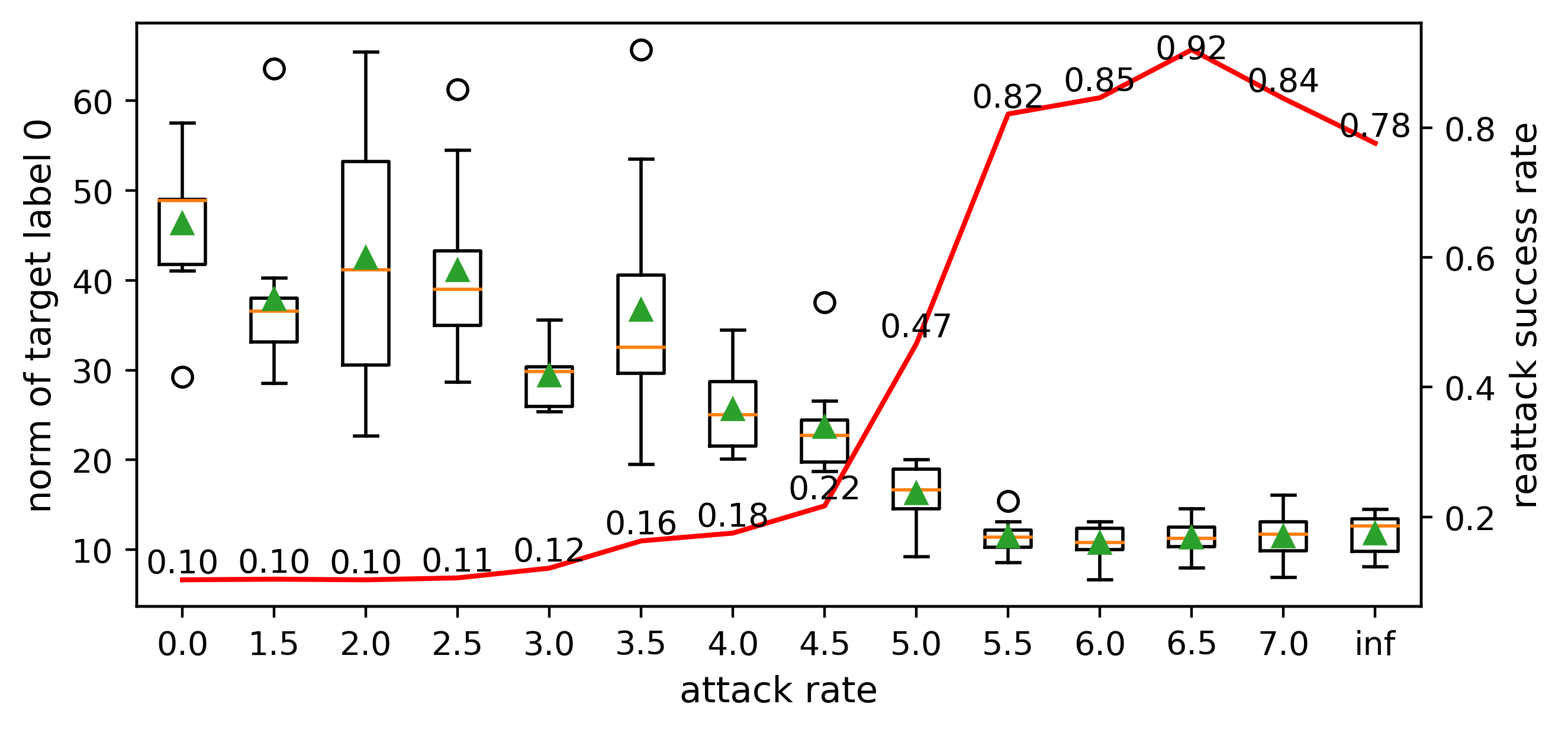}
        \captionsetup{skip=0pt}
        \caption{neural cleanse results (lambda=0.01)}
        \label{fig:nc-0.01}
    \end{subfigure}
    
    \caption{The impacts of different attack rates on Neural Cleanse. $ar=0.0$ and $ar=inf$ represents the results of benign models and baseline backdoored models, respectively. The left y-axis represents the norm of the reversed triggers, which is represented by the box plots. The right y-axis represents the reattack success rate, which is represented by the curve.}
    \label{fig:ar-nc}
    \vspace{-0.3cm}
\end{figure}

\begin{table}[tb]
    \small
    \centering
    \begin{tabular}{c|cccc}
        \toprule
        Attack Rate  & 2.0 & 3.0 & 4.0 & 4.5 \\
        \midrule
        Confidence   & 23.20\% & 45.09\% & 69.06\% & 78.63\% \\
        CE Loss & 1.4612 & 0.7966 & 0.3702 & 0.2404 \\
        \midrule
        Attack Rate  & 5.0 & 6.0 & 6.5 & 7.0\\
        \midrule
        Confidence   & 85.85\% & 94.28\% & 96.45\% & 97.82\% \\
        CE Loss & 0.1526 & 0.0589 & 0.0361 & 0.0221 \\
        \bottomrule
    \end{tabular}
    \caption{The variations of classification confidence and cross entropy loss, as attack rate changes in 10-class classification tasks.}
    \label{tab:ar-details}
    \vspace{-0.3cm}
\end{table}

Fig. \ref{fig:ar-nc} shows the defense performance of Neural Cleanse when attack rate changes. We fix $\lambda$ in Neural Cleanse to 0.001 and 0.01 in this experiment, for better illustrations. Note that Neural Cleanse further check the MAD value to determine whether a model is backdoored. Hence, Fig. \ref{fig:ar-nc} can only show the performance of reverse engineering and it is not equivalent to the detection performance. To verify the validity of reversed triggers, we introduce a new metric called reattack success rate (ReASR). Let $m_{gt}$ be the ground truth mask of backdoor trigger, $m'$ be the reversed mask, and $\Delta'$ be the reversed trigger pattern. We calculate the attack success rate of $m_{gt}\cdot m'\cdot \Delta'$ as ReASR. This metric can eliminate the influence of universal adversarial perturbation and directly reveal the similarity between the reversed trigger and the ground truth trigger. From Fig. \ref{fig:nc-0.001} and Fig. \ref{fig:nc-0.01}, we can both observe a bound value in ReASR (e.g., $ar=6.0$ in Fig. \ref{fig:nc-0.001} and $ar=4.5$ in Fig. \ref{fig:nc-0.01}). Below these bound values, the reversed triggers are basically ineffective in inducing backdoor behavior. However, when $ar$ is larger than these bound values, ReASR increase rapidly. This phenomenon can be well explained by our compensatory model proposed in Sec. \ref{sec:theory}. In Fig. \ref{fig:nc-0.001}, the norm of triggers of the benign model is 50.06, and that of the baseline backdoored model is 14.28, thus $\lambda \cdot \left ( \| m^{ben} \| - \| m^{poi}\| \right)$ is 0.0358. According to Tab. \ref{tab:ar-details}, when $ar \leq 6.0$, the cross entropy loss is much larger than $\lambda \cdot \left ( \| m^{ben}\| - \| m^{poi}\| \right)$, which satisfies the constraint in Eq. \eqref{attack neural cleanse}. Thus, Neural Cleanse cannot reconstruct the trigger when $ar \leq 6.0$. When $ar \geq 6.5$, the effect of Eq. \eqref{attack neural cleanse} weakens, which allows Neural Cleanse to partially reconstruct the trigger. For Fig. \ref{fig:nc-0.01}, the value of $\lambda \cdot \left ( \| m^{ben}\| - \| m^{poi}\| \right)$ is 0.3449, which leads to a bound around $ar=4.5$. Due to space limitations, the analysis on ABS is provided in the supplementary materials.

\section{Summary and Ethical Statement}

In this paper, we analyzed the existing reversing based backdoor defense methods, and constructed a generic paradigm for the trigger reverse engineering process. Based on the paradigm, we proposed a new attack perspective to defeat the reversing based methods, via manipulating the classification confidence of backdoor samples. To quantify the manipulations, we proposed a compensatory model to compute the lower bound of the modifications for effective attacks. Based on the compensatory model, we proposed a plug-and-play attack framework, named LSP framework, which can be effectively integrated with the existing backdoor attacks. Extensive experiments demonstrated that our LSP framework is compatible with various existing backdoor attacks and can easily disable the reversing based defense methods. We believe that exploring the design flaws in reversing based methods can help to improve our understanding about the defense mechanisms and thus accelerate the development of backdoor defense techniques.

\appendix

\bibliographystyle{named}
\bibliography{ijcai24}

\begin{thebibliography}{}

\bibitem[\protect\citeauthoryear{Bai \bgroup \em et al.\egroup }{2021}]{DBLP:conf/iclr/BaiWZL0X21}
Jiawang Bai, Baoyuan Wu, Yong Zhang, Yiming Li, Zhifeng Li, and Shu{-}Tao Xia.
\newblock Targeted attack against deep neural networks via flipping limited weight bits.
\newblock In {\em 9th International Conference on Learning Representations}, 2021.

\bibitem[\protect\citeauthoryear{Brown \bgroup \em et al.\egroup }{2017}]{DBLP:journals/corr/abs-1712-09665}
Tom~B. Brown, Dandelion Man{\'{e}}, Aurko Roy, Mart{\'{\i}}n Abadi, and Justin Gilmer.
\newblock Adversarial patch.
\newblock {\em CoRR}, abs/1712.09665, 2017.

\bibitem[\protect\citeauthoryear{Chen \bgroup \em et al.\egroup }{2017}]{DBLP:journals/corr/abs-1712-05526}
Xinyun Chen, Chang Liu, Bo~Li, Kimberly Lu, and Dawn Song.
\newblock Targeted backdoor attacks on deep learning systems using data poisoning.
\newblock {\em CoRR}, abs/1712.05526, 2017.

\bibitem[\protect\citeauthoryear{Chen \bgroup \em et al.\egroup }{2019}]{DBLP:conf/aaai/ChenCBLELMS19}
Bryant Chen, Wilka Carvalho, Nathalie Baracaldo, Heiko Ludwig, Benjamin Edwards, Taesung Lee, Ian~M. Molloy, and Biplav Srivastava.
\newblock Detecting backdoor attacks on deep neural networks by activation clustering.
\newblock In {\em In proceedings of the Workshop on Artificial Intelligence Safety}, 2019.

\bibitem[\protect\citeauthoryear{Gu \bgroup \em et al.\egroup }{2019}]{DBLP:journals/access/GuLDG19}
Tianyu Gu, Kang Liu, Brendan Dolan{-}Gavitt, and Siddharth Garg.
\newblock Badnets: Evaluating backdooring attacks on deep neural networks.
\newblock {\em {IEEE} Access}, 7:47230--47244, 2019.

\bibitem[\protect\citeauthoryear{Ho{-}Phuoc}{2018}]{DBLP:journals/corr/abs-1811-07270}
Tien Ho{-}Phuoc.
\newblock {CIFAR10} to compare visual recognition performance between deep neural networks and humans.
\newblock {\em CoRR}, abs/1811.07270, 2018.

\bibitem[\protect\citeauthoryear{Hu \bgroup \em et al.\egroup }{2023}]{hu2023planning}
Yihan Hu, Jiazhi Yang, Li~Chen, Keyu Li, Chonghao Sima, Xizhou Zhu, Siqi Chai, Senyao Du, Tianwei Lin, Wenhai Wang, et~al.
\newblock Planning-oriented autonomous driving.
\newblock In {\em Proceedings of the IEEE/CVF Conference on Computer Vision and Pattern Recognition}, 2023.

\bibitem[\protect\citeauthoryear{Huang \bgroup \em et al.\egroup }{2022}]{DBLP:conf/iclr/Huang0WQ022}
Kunzhe Huang, Yiming Li, Baoyuan Wu, Zhan Qin, and Kui Ren.
\newblock Backdoor defense via decoupling the training process.
\newblock In {\em The Tenth International Conference on Learning Representations}, 2022.

\bibitem[\protect\citeauthoryear{Jiang \bgroup \em et al.\egroup }{2023}]{DBLP:conf/cvpr/Jiang0X023}
Wenbo Jiang, Hongwei Li, Guowen Xu, and Tianwei Zhang.
\newblock Color backdoor: {A} robust poisoning attack in color space.
\newblock In {\em {IEEE/CVF} Conference on Computer Vision and Pattern Recognition}, 2023.

\bibitem[\protect\citeauthoryear{Le and Yang}{2015}]{le2015tiny}
Ya~Le and Xuan Yang.
\newblock Tiny imagenet visual recognition challenge.
\newblock {\em CS 231N}, 7(7):3, 2015.

\bibitem[\protect\citeauthoryear{Lin \bgroup \em et al.\egroup }{2020}]{DBLP:conf/ccs/LinXL020}
Junyu Lin, Lei Xu, Yingqi Liu, and Xiangyu Zhang.
\newblock Composite backdoor attack for deep neural network by mixing existing benign features.
\newblock In {\em {CCS} '20: 2020 {ACM} {SIGSAC} Conference on Computer and Communications Security}, 2020.

\bibitem[\protect\citeauthoryear{Liu \bgroup \em et al.\egroup }{2018a}]{DBLP:conf/raid/0017DG18}
Kang Liu, Brendan Dolan{-}Gavitt, and Siddharth Garg.
\newblock Fine-pruning: Defending against backdooring attacks on deep neural networks.
\newblock In {\em Research in Attacks, Intrusions, and Defenses}, 2018.

\bibitem[\protect\citeauthoryear{Liu \bgroup \em et al.\egroup }{2018b}]{DBLP:conf/ndss/LiuMALZW018}
Yingqi Liu, Shiqing Ma, Yousra Aafer, Wen{-}Chuan Lee, Juan Zhai, Weihang Wang, and Xiangyu Zhang.
\newblock Trojaning attack on neural networks.
\newblock In {\em 25th Annual Network and Distributed System Security Symposium}, 2018.

\bibitem[\protect\citeauthoryear{Liu \bgroup \em et al.\egroup }{2019}]{DBLP:conf/ccs/LiuLTMAZ19}
Yingqi Liu, Wen{-}Chuan Lee, Guanhong Tao, Shiqing Ma, Yousra Aafer, and Xiangyu Zhang.
\newblock {ABS:} scanning neural networks for back-doors by artificial brain stimulation.
\newblock In {\em Proceedings of the 2019 {ACM} {SIGSAC} Conference on Computer and Communications Security}, 2019.

\bibitem[\protect\citeauthoryear{Liu \bgroup \em et al.\egroup }{2022}]{DBLP:conf/cvpr/LiuSTWM022}
Yingqi Liu, Guangyu Shen, Guanhong Tao, Zhenting Wang, Shiqing Ma, and Xiangyu Zhang.
\newblock Complex backdoor detection by symmetric feature differencing.
\newblock In {\em {IEEE/CVF} Conference on Computer Vision and Pattern Recognition}, 2022.

\bibitem[\protect\citeauthoryear{Marriott \bgroup \em et al.\egroup }{2021}]{marriott20213d}
Richard~T Marriott, Sami Romdhani, and Liming Chen.
\newblock A 3d gan for improved large-pose facial recognition.
\newblock In {\em Proceedings of the IEEE/CVF Conference on Computer Vision and Pattern Recognition}, 2021.

\bibitem[\protect\citeauthoryear{Nguyen and Tran}{2020}]{DBLP:conf/nips/NguyenT20}
Tuan~Anh Nguyen and Anh~Tuan Tran.
\newblock Input-aware dynamic backdoor attack.
\newblock In {\em Advances in Neural Information Processing Systems 33}, 2020.

\bibitem[\protect\citeauthoryear{Nguyen and Tran}{2021}]{DBLP:conf/iclr/NguyenT21}
Tuan~Anh Nguyen and Anh~Tuan Tran.
\newblock Wanet - imperceptible warping-based backdoor attack.
\newblock In {\em 9th International Conference on Learning Representations}, 2021.

\bibitem[\protect\citeauthoryear{Qiu \bgroup \em et al.\egroup }{2021}]{DBLP:conf/asiaccs/0001ZGZQT21}
Han Qiu, Yi~Zeng, Shangwei Guo, Tianwei Zhang, Meikang Qiu, and Bhavani Thuraisingham.
\newblock Deepsweep: An evaluation framework for mitigating {DNN} backdoor attacks using data augmentation.
\newblock In {\em {ASIA} {CCS} '21: {ACM} Asia Conference on Computer and Communications Security}, 2021.

\bibitem[\protect\citeauthoryear{Stallkamp \bgroup \em et al.\egroup }{2012}]{DBLP:journals/nn/StallkampSSI12}
Johannes Stallkamp, Marc Schlipsing, Jan Salmen, and Christian Igel.
\newblock Man vs. computer: Benchmarking machine learning algorithms for traffic sign recognition.
\newblock {\em Neural Networks}, 32:323--332, 2012.

\bibitem[\protect\citeauthoryear{Szegedy \bgroup \em et al.\egroup }{2016}]{DBLP:conf/cvpr/SzegedyVISW16}
Christian Szegedy, Vincent Vanhoucke, Sergey Ioffe, Jonathon Shlens, and Zbigniew Wojna.
\newblock Rethinking the inception architecture for computer vision.
\newblock In {\em 2016 {IEEE} Conference on Computer Vision and Pattern Recognition}, 2016.

\bibitem[\protect\citeauthoryear{Turner \bgroup \em et al.\egroup }{2019}]{DBLP:journals/corr/abs-1912-02771}
Alexander Turner, Dimitris Tsipras, and Aleksander Madry.
\newblock Label-consistent backdoor attacks.
\newblock {\em CoRR}, abs/1912.02771, 2019.

\bibitem[\protect\citeauthoryear{Wang \bgroup \em et al.\egroup }{2004}]{DBLP:journals/tip/WangBSS04}
Zhou Wang, Alan~C. Bovik, Hamid~R. Sheikh, and Eero~P. Simoncelli.
\newblock Image quality assessment: from error visibility to structural similarity.
\newblock {\em {IEEE} Trans. Image Process.}, 13(4):600--612, 2004.

\bibitem[\protect\citeauthoryear{Wang \bgroup \em et al.\egroup }{2019}]{DBLP:conf/sp/WangYSLVZZ19}
Bolun Wang, Yuanshun Yao, Shawn Shan, Huiying Li, Bimal Viswanath, Haitao Zheng, and Ben~Y. Zhao.
\newblock Neural cleanse: Identifying and mitigating backdoor attacks in neural networks.
\newblock In {\em 2019 {IEEE} Symposium on Security and Privacy}, 2019.

\bibitem[\protect\citeauthoryear{Wenger \bgroup \em et al.\egroup }{2021}]{DBLP:conf/cvpr/WengerPBY0Z21}
Emily Wenger, Josephine Passananti, Arjun~Nitin Bhagoji, Yuanshun Yao, Haitao Zheng, and Ben~Y. Zhao.
\newblock Backdoor attacks against deep learning systems in the physical world.
\newblock In {\em {IEEE} Conference on Computer Vision and Pattern Recognition}, 2021.

\bibitem[\protect\citeauthoryear{Xiao \bgroup \em et al.\egroup }{2017}]{DBLP:journals/corr/abs-1708-07747}
Han Xiao, Kashif Rasul, and Roland Vollgraf.
\newblock Fashion-mnist: a novel image dataset for benchmarking machine learning algorithms.
\newblock {\em CoRR}, abs/1708.07747, 2017.

\bibitem[\protect\citeauthoryear{Zhao \bgroup \em et al.\egroup }{2022}]{DBLP:conf/cvpr/Zhao0XDWL22}
Zhendong Zhao, Xiaojun Chen, Yuexin Xuan, Ye~Dong, Dakui Wang, and Kaitai Liang.
\newblock {DEFEAT:} deep hidden feature backdoor attacks by imperceptible perturbation and latent representation constraints.
\newblock In {\em {IEEE/CVF} Conference on Computer Vision and Pattern Recognition}, 2022.

\end{thebibliography}

\end{document}